# Limitations of Rotational Acceleration for Predicting Concussion: A New Injury Risk Function

Jessica A. Towns[1], Nicholas J. Cecchi[1,2], James W. Hickey[3], William T. O'Brien[3], Spencer S.H. Roberts[4], Gerald A. Grant[5], Michael M. Zeineh[6], Stuart J. McDonald[3,7], David B. Camarillo[1,2*]

[1]Department of Bioengineering, Stanford University, Stanford, 94305, USA

*Corresponding author. Email: dcamarillo@stanford.edu

[2]SoftShox, Vienna, VA, USA

[3]Department of Neuroscience, School of Translational Medicine, Monash University, Melbourne, 3800, Australia

[4]Centre for Sport Research, Institute for Physical Activity and Nutrition, Deakin University, Geelong, 3220, Australia

[5]Department of Neurosurgery, Duke University, Durham, 27708, USA

[6]Department of Radiology, Stanford University, Stanford, 94305, USA

[7]Department of Neurology, Alfred Health, Melbourne, 3004, Australia

# Abstract

**Purpose:** Rotational acceleration of the head has long been hypothesized to be a primary driver of concussion because it can generate shear strains in brain tissue, while linear acceleration has been considered less injurious due to the brain's near-incompressibility. Instrumented mouthguards (iMGs) now enable direct six-degree-of-freedom measurements of human concussion, making it possible to rigorously test this hypothesis for the first time *in vivo*. The purpose of this study was to test the hypothesis that rotational acceleration is a more precise predictor of concussion than linear acceleration using iMG measurements of concussion across a multi-sport dataset.

**Methods:** A dataset of 47 concussive and 3,805 non-concussive iMG-measured impacts was compiled across male and female athletes in seven sports. Univariate and bivariate logistic regression models were developed using peak linear acceleration, peak rotational acceleration, and peak rotational velocity to assess concussion classification and contribution to injury risk. Models were evaluated using area under the precision-recall curve (AUPRC), F1 score, and odds ratios. Additionally, a novel liquid-filled helmet technology was evaluated during laboratory impacts using the derived risk functions.

**Results:** Contrary to our hypothesis, rotational acceleration was not a significantly more precise classifier than linear acceleration (AUPRC=0.35 vs 0.65; F1=0.31 vs 0.50), and was not a significant independent predictor of concussion alongside linear acceleration in a bivariate model ($OR_{rot}$=1.3, 95% CI: 0.81-1.9; $OR_{lin}$=2.3; 95% CI: 1.5-3.6). However, peak rotational velocity provided additional predictive value in a bivariate risk function with peak linear acceleration. Applying this bivariate function to laboratory football helmet tests, the liquid-filled padding reduced predicted concussion risk by up to 52% compared to a standard American football helmet, primarily due to attenuating linear acceleration.

**Conclusion:** Rotational acceleration was not a significant independent predictor of concussion, and was outperformed by linear acceleration in classification, challenging its primacy in injury risk evaluation. Our results support a broader consideration of linear acceleration and rotational velocity in concussion risk assessment and prevention.

## Introduction

Traumatic brain injury (TBI) is the second most common neurological disorder in the United States,[1] with an estimated global incidence of nearly 69 million cases per year.[2] Approximately 80−90% of TBIs are classified as "mild" (mTBI or concussion),[3] yet they can still result in persistent post-concussive sequelae one year after the injury, including deficits in attention, memory, and executive function,[4–7] as well as psychiatric comorbidities which may persist for years.[8–11] The biomechanical cause of closed-head TBI was first proposed during World War II, when A.H.S. Holbourn hypothesized that rotational acceleration of the head induces large shear strains within brain tissue, whereas linear acceleration does not.[12] The biophysical basis for the rotational acceleration hypothesis is that the massive human brain is a soft, gel-like structure that shears as it lags behind the rotating skull.[13] In contrast to rotational motion, linear acceleration primarily transmits compressive forces to the head rather than shear. Because the brain and cerebrospinal fluid are nearly incompressible, compressive or tensile loading were supposed to cause little deformation of the brain. As a result, rotational head kinematics and brain strain have become the primary diffuse brain injury criteria for many concussion safety testing methodologies.

Early support for the rotational hypothesis emerged from large animal models of moderate-to-severe TBI, and later shifted toward mTBI, or concussion. During the 1960s and 70s, Ommaya and Gennarelli demonstrated that rotational loading could produce loss of consciousness and widespread axonal pathology in non-human primates.[14–16] Others have since extended this work using porcine models, applying rapid rotational accelerations to replicate milder injury patterns observed in human concussion.[17] Finite element modelling of the porcine brain has since shown that rotational acceleration induces diffuse strain in the deep white matter,[18] a region heavily implicated pathologically.[19,20] Together, these studies motivated the need for human measurements to translate findings from animals and establish clinically relevant concussion thresholds.

With the miniaturization of electromechanical systems (MEMS), it became possible to measure head impact kinematics with wearable sensors on humans and move toward developing injury risk functions. Injury risk functions estimate the probability of brain injury from mechanical inputs such as head motion, and are used to establish injury thresholds and evaluate protective equipment performance. Beginning in the early 2000s, the Head Impact Telemetry System (HITS)[21] enabled the first large-scale data collection studies in helmeted sports by embedding accelerometers into American football helmets. Since then, HITS data has informed the Brain Injury Criterion, a rotational velocity-based injury risk function[22] proposed by the National Highway Traffic Safety Administration for automotive crash testing, and the Summation of Tests for the Analysis of Risk (STAR) helmet rating system.[23] HITS used an array of helmet-mounted accelerometers to measure linear acceleration and estimated rotational acceleration using a centroid-based algorithm.[24] However, it did not capture time-resolved rotational acceleration and was affected by relative motion between the helmet and skull, reducing accuracy.[24,25]

Instrumented mouthguards (iMGs) overcome limitations of motion artifacts by rigidly coupling inertial sensors to the upper dentition, enabling direct six-degree-of-freedom measurements of linear and rotational head kinematics *in vivo.*[26] IMGs have been extensively tested in laboratory and on-field settings,[27–29] and deployed at scale in both helmeted and unhelmeted sports,[30,31] capturing large head impact datasets under real-world conditions. With these data now available, iMG technology enables rigorous testing of the rotational acceleration hypothesis for the first time in humans. In the current study, we used a dataset of iMG-measured head impacts to test the long-standing hypothesis that rotational acceleration is a more precise predictor of concussion than linear acceleration. Surprisingly, we show that that linear acceleration is a significantly more precise predictor of concussion than rotational acceleration, and propose two risk functions incorporating linear acceleration. We further demonstrate that concussion risk remains substantially high during laboratory impacts to American football helmets and that attenuating linear acceleration with a liquid-pad helmet design can meaningfully reduce predicted concussion risk.

# Methods

Mouthguard data was collated among cohorts from American football, Australian football, lacrosse, rugby union, artistic gymnastics, mixed martial arts, and ice hockey, where mouthguard sensors recorded linear and rotational head kinematics. Concussions were diagnosed by trained personnel and confirmed by video review. Video-verified impact kinematics were used to develop injury risk functions and estimate tissue-based criteria using finite element analysis. Injury prediction was investigated using univariate and bivariate logistic regression. Finally, a liquid-based helmet technology was evaluated during laboratory helmet impacts using the best-performing bivariate risk function.

### *Data description*

A total of 3,805 non-concussive and 47 concussive impacts were compiled to construct the dataset for concussion risk modeling and classification (Table 1). Concussions were either medically diagnosed or labeled based on visual signs of concussion observed during video review.[32] Kinematic data were collected with custom-fitted instrumented mouthguards, including the Prevent iMG,[32] the In-mouth ADXL/L3G4200D,[33] the Stanford MiG2.0,[29] the HITIQ Smart Mouthguard,[27] and the Wake Forest mouthpiece.[34] The dataset was sourced from youth ice hockey,[35] collegiate women's lacrosse, professional mixed martial arts (MMA),[36] collegiate rugby union, women's advanced gymnastics,[37] amateur Australian football, and youth, high-school, and collegiate American football.[32,38] Among concussive impacts, 17 occurred in American football, 19 in Australian football, 5 in MMA, 3 in rugby union, 1 in gymnastics, 1 in ice hockey, and 1 in lacrosse. Three concussive impacts involved female athletes in lacrosse, rugby, and gymnastics. Youth was defined as below 14 years of age.

### *Concussion diagnosis and impact identification*

Given the collation of multiple published datasets, concussion diagnosis procedures differed slightly across studies. To ensure consistency, we harmonized concussion identification using two criteria: (1) the presence of visual signs or reported symptoms of injury and (2) visual confirmation of the associated head impact. Importantly, the clinical signs and symptoms used to diagnose concussion are internationally harmonized through successive Concussion in Sport Group consensus statements, which outline a common set of diagnostic indicators across continents and sports, such as disorientation, motor incoordination, balance disturbance, and athlete-reported symptoms, among many others.[39,40] This global standardization supports the comparability of concussion cases across the American and Australian datasets included in the current study.

In studies with on-site medical staff (youth American football, ice hockey, gymnastics, lacrosse, rugby, Australian football, and Stanford football) concussions were immediately diagnosed by a medical professional based on acute signs, and video review was performed to isolate the associated impact. In the MMA study, post-bout clinical evaluations were performed by a medical professional, which included the neurocognitive SCAT5 assessment Because multiple strikes occur in rapid succession without opportunities for in-bout medical assessment, the concussive impact was defined as the largest rotational acceleration impact that occurred during the bout. The chosen impact was subsequently verified by video review. Finally, in the Bartsch et al. American football study, both diagnosis and impact identification were performed through retrospective video review for visual signs of concussion.

### *Australian football, rugby union and lacrosse field data collection*

Data collection methods for the published datasets included in our analysis are described in prior work (Table 1), while collection methods for unpublished datasets are described here. The Australian football data builds on a 2022 two-team pilot study within the Victorian Amateur Football Association (VAFA) by Evans et al[31] and is primarily drawn from a broader VAFA cohort study investigating blood-based biomarkers of brain injury.[41] A subset of 11 men's teams (186 players) wore HITIQ iMGs over the 2022–2024 seasons to capture impacts exceeding an 8 g threshold. Video verification was conducted using the same procedures,[31] applied to all timestamped concussion cases and a subset of non-concussive impacts from selected matches, based on events passing HITIQ's filtering pipeline; data from verified cases were then reprocessed using a 200 Hz filter. Australian football projects were approved by the Monash University Human Research Ethics Committee (Project IDs: 27684, 36297, 36556) and Deakin University (Project ID: 2022-337).

| **Authors** | **Sport** | **Level(s)** | **Gender(s)** | **Measurement Device** | **No. Concussions** | **No. Non-concussions** | **Diagnostic criteria** |
|---|---|---|---|---|---|---|---|
| Current study | Lacrosse | Collegiate | Female | Stanford MiG2.0 | 1 | 102 | Immediate diagnosis |
| Current study | Rugby | Collegiate | Female/ Male | Prevent iMG | 3 | 469 | Immediate diagnosis/ player identification of impact |
| Domel et al.[42] | American football | Collegiate | Male | Stanford MiG2.0 | 0 | 359 | |
| Hernandez et al.[38] | American football | Collegiate | Male | In-Mouth (ADXL377/ L3G4200D) | 1 | 0 | Immediate diagnosis |
| Bartsch et al.[32] | American football | High-school, collegiate | Male | Prevent iMG | 13 | 1782 | Visual signs of concussion |
| Tiernan et al.[36] | Mixed martial arts | Professional, semi-professional, amateur | Male | Stanford MiG 2.0 | 5 | 304 | Largest acceleration impact immediately prior to diagnosis |
| Evans et al.[31] and ongoing study | Australian football | Amateur | Male | HITIQ | 19 | 776 | Immediate diagnosis |
| NIH Grant K25HD101686 | American football | Youth | Male | WF | 3 | 0 | Immediate diagnosis |
| Pritchard et al.[37] | Gymnastics | Advanced | Female | WF | 1 | 0 | Immediate diagnosis |
| Swenson et al.[35] | Ice hockey | Youth | Male | WF | 1 | 0 | Immediate diagnosis |

**Table 1.** Head impact datasets used in this study.

Additionally, as part of a larger ongoing study, the Prevent iMG was deployed to 21 collegiate rugby players (7 male, 14 female) on the Stanford rugby team to collect data for the complete 2022 season, while the Stanford MiG2.0 was deployed to 11 collegiate women's lacrosse players across 6 practices in the Spring 2018 season. All mouthguards were custom-fitted to the athlete's upper dentition and equipped with a triaxial accelerometer and triaxial gyroscope to record linear accelerations and rotational velocities during head impacts (Fig. 1a). Video footage was recorded at 4k resolution and 60 frames per second for all athlete-exposures using two camera angles to capture both halves of the field. A world clock was filmed at the start of the recording session to enable temporal synchronization between mouthguard-recorded events and session footage. All impacts exceeding a 10g trigger threshold were subject to visual verification, and only video-confirmed impacts were included in our analyses. Data collection was approved by the Stanford University Institutional Review Board (Protocol 45932).

### *Mouthguard kinematics processing*

Kinematics for video-verified impacts measured with the Prevent, HITIQ, Stanford, and In-Mouth iMGs were first transformed to the SAE J211 coordinate system[43] (+X: forward, +Y, to the right, +Z downward). Linear acceleration and rotational velocity were then filtered using a fourth-order Butterworth low pass phaseless filter with a 200 Hz cutoff frequency (-6 dB). Rotational accelerations were next derived from rotational velocity measurements using a five-point stencil derivative and subsequently filtered at 200 Hz. Finally, linear accelerations were transformed to the head's center of gravity using the respective sensor locations for each mouthguard and Equation 1,

$$a^{CG} = a^{MG} + \alpha^{MG} \times \vec{r} + \omega^{MG} \times (\omega^{MG} \times \vec{r}) \quad (1)$$

where $a^{CG}$ is the linear acceleration (m/s$^2$) at the head's center of gravity for a 50$^{th}$ percentile male, $a^{MG}$ is the linear acceleration of the mouthguard, $\alpha$ is rotational acceleration (rad/s$^2$), $\omega$ is rotational velocity (rad/s), and $\vec{r}$ is the constant distance vector from the accelerometer to the head's center of gravity. The 200 Hz frequency was chosen due to the unavailability of the raw kinematic data measured from the Prevent iMG in the American football dataset, which comprises 47% of our data. The sampling frequency of the iMGs used in this study are large enough such that a 200 Hz frequency cutoff avoids signal aliasing. Methods for evaluating measurement consistency across devices after 200 Hz filtering are provided in the Supplementary Information, and outcomes are shown in Supplementary Table 6 and Supplementary Figure 1. IMG specifications are also provided in Supplementary Table 5.

Impacts that were measured by the Wake Forest mouthpiece followed a slightly different approach. First, data were filtered using a fourth-order Butterworth low pass phaseless filter with a 1,650 Hz cutoff frequency (CFC1000) for linear acceleration and a 250 Hz cutoff frequency for rotational velocity. The impact measured from the gymnastics participant was not initially filtered due to the lower 350 Hz sampling rate (Supplementary

Table 5). The accelerometer baseline was zero-offset by subtracting the first data point from all subsequent samples. The gyroscope baseline was zero-offset by subtracting the fifth data point from all subsequent samples and setting the preceding 4 samples to zero. Data were then rotated to a conventional coordinate system (+X: forward, +Y, to the left, +Z upward). Rotational acceleration was derived from rotational velocity using the five-point stencil derivative method. Linear acceleration was then transformed to the head's center of gravity using Equation 1. For inclusion in this study, secondary filtering was applied with a 200 Hz cutoff frequency and data were re-rotated to follow the SAEJ211 coordinate system. For all iMGs, peak kinematic values were calculated from the resultant, or Euclidian norm, of the x, y and z directions.

Head impact power (HIP) is defined as the rate at which rotational and translational kinetic energy is transferred to the head, and was calculated using Equation 2,

$$HIP(t) = m \cdot a_x \int a_x dt + m \cdot a_y \int a_y dt + m \cdot a_z \int a_z dt + I_{xx} \cdot \alpha_x \cdot \omega_x + I_{yy} \cdot \alpha_y \cdot \omega_y + I_{zz} \cdot \alpha_z \cdot \omega_z \tag{2}$$

where m is head mass (kg), $I_{xx}$, $I_{yy}$, and $I_{zz}$ are the directional components of the head's moment of inertia (kg·cm$^2$), a is linear acceleration (m/s$^2$), α is rotational acceleration (rad/s$^2$), and ω is rotational velocity (rad/s), further split into directional components. We used an average head mass (m) of 4.1 kg and 3.2 kg for male and female athletes, respectively, based on cadaveric data.[44,45] The moments of inertia are described by equations 3-5.[44,45]

$$I_{xx}(kg \cdot cm^2) = 74.8m - 125.5 \tag{3}$$

$$I_{yy}(kg \cdot cm^2) = 72.4m - 90.2 \tag{4}$$

$$I_{zz}(kg \cdot cm^2) = 45.6m - 26.5 \tag{5}$$

### *Mouthguard data cleaning*

All impacts included in this study underwent video verification to ensure impact timestamps aligned with visual events. For MMA, the concussive impact was defined as the impact with the highest rotational acceleration recorded during the competitive event in which the concussion was diagnosed.[36] This method was used because head impacts occur frequently during competition, making it difficult to isolate a single concussive event when multiple potentially injurious impacts may occur in the same competitive bout. Additionally, one Australian football concussion was excluded since the infrared proximity sensor reading fell below the threshold used to classify true impacts and was therefore considered a false negative. Low proximity readings may reflect decoupling or recoupling between the mouthguard and upper dentition, which can produce noisy fluctuations or unusually low magnitudes relative to the observed head motion.[15] Therefore, this Australian football concussion was removed.

### *Finite element modeling of head impacts*

Maximum principal strains were computed using the Global Human Body Model Consortium 50$^{th}$ percentile male head and brain model (GHBMC M50-H v6.1). Brain tissue responses from the GHBMC model were previously validated with cadaveric data.[46] Mouthguard data were transformed to align with the SAEJ211 coordinate system of the GHBMC model. Linear accelerations were applied to the head's center of gravity and rotational velocities were applied globally. The 95$^{th}$ percentile maximum principal strains (MPS95) were extracted regionally and globally using a custom MATLAB script. Considered brain regions included the corpus callosum, brainstem, midbrain, basal ganglia, cerebral white matter, upper gray matter, lower gray matter, thalamus and cerebellum.

### *Statistical methods*

#### *Kinematics and strain comparisons*

All statistical analyses and model development were performed in RStudio (2024.09.0+375). Differences in peak kinematics and MPS95 between concussive and non-concussive impacts were tested using a Wilcoxon Rank Sum test with a Bonferroni correction ($n=7$, $p<0.007$). Additionally, to evaluate consistency across sensor measurements after filtering, empirical cumulative distribution functions of peak linear acceleration, peak rotational velocity, and their primary frequencies were compared between iMGs using a Kolmogorov-Smirnov test (Supplementary Figure 1 and Supplementary Table 6). Further details on the statistical comparisons are described in the Supplementary Information.

#### *Logistic regression models*

We employed logistic regression models in order to 1) quantify how biomechanical metrics relevant to sports and automotive safety classify concussion and 2) derive injury risk functions. We first divided the full dataset into an 80% training set (n=3,078) and 20% held-out test set (n=744) with a fixed random seed and stratified sampling to preserve concussion incidence. Additionally, youth concussions from the Wake Forest cohorts were withheld as a separate test set. All model threshold tuning described below were performed on the training set alone.

On the training set, rare-events logistic models were fit using ReLogit to address small-sample bias inherent in maximum likelihood estimation when the outcome of interest is rare.[47] Twelve candidate models were tested, including peak linear acceleration ($\mathbf{a}$), peak rotational acceleration ($\boldsymbol{\alpha}$), peak rotational velocity ($\boldsymbol{\omega}$), $\mathbf{a} + \boldsymbol{\alpha}$, $\mathbf{a} + \boldsymbol{\omega}$, peak total head impact power (HIP),[48] peak rotational power ($HIP_{rot}$), peak linear power ($HIP_{lin}$), Head Injury Criterion ($HIC_{36}$),[49] Brain Injury Criterion (BrIC),[22] Combined Probability (CP),[50] and MPS95. Further justification on the candidate models is described in the Supplementary Information. A prior correction of 5.5 concussions per 1,000 impacts was applied to the intercept term, which is the assumed concussion rate in collegiate American football.[51] This correction reduces bias in baseline probability estimates without affecting the slope coefficients. Models were implemented using the Zelig() function, with standard errors clustered by subject ID to account for within-player correlation. Predictors were standardized to have a mean of 0 and standard deviation of 1. Model fit was assessed by deviance ($G^2$) and Bayesian Information Criterion.

Classification thresholds were optimized on the training set by sweeping from 0 to 1 in increments of 0.1 to maximize F1. Determined thresholds were then applied to the 20% held-out test set and performance metrics—including area under the precision−recall curve (AUPRC), F1, precision, and recall—were estimated with 95% confidence intervals using a stratified bootstrap approach over 1,000 iterations. The same thresholds were applied to the held-out youth concussions (n=4) and recall (sensitivity) was calculated.

*Injury Risk Functions*

To determine which logistic regression models to analyze as injury risk functions and subject to statistical comparison, we performed an initial screening using deviance ($G^2$), Bayesian Information Criterion, LASSO-penalized logistic regression, and dominance analysis. $G^2$ and Bayesian Information Criterion were used to assess model fit, while LASSO and dominance analysis evaluated relative predictor importance based on coefficient magnitude and variance explained, respectively. Predictors or models that ranked highly across these approaches were selected for injury risk modeling and statistical comparison. Full methodological details on the LASSO and dominance analyses are provided in the Supplementary Information.

The final subset of features selected for injury risk function development included $\mathbf{a}$, $\boldsymbol{\omega}$ , $\boldsymbol{\alpha}$, and $\boldsymbol{\alpha}$ based on consistently ranking as the top three predictors across both LASSO regression and dominance analysis. These three features also demonstrated the strongest model fit ($G^2$ and BIC) relative to other predictor variables. Therefore, univariate injury risk functions included $\mathbf{a}$, $\boldsymbol{\alpha}$, and $\boldsymbol{\omega}$, while bivariate functions included linear combinations of $\mathbf{a}$ and $\boldsymbol{\omega}$ (Eq. 6) and $\mathbf{a}$ and $\boldsymbol{\alpha}$ (Eq. 7),

$$P_{inj}(a + \omega) = \frac{1}{1 + e^{[-(\beta_0 + \beta_1 a + \beta_2 \omega)]}} \quad (6)$$

$$P_{inj}(a + \alpha) = \frac{1}{1 + e^{[-(\beta_0 + \beta_1 a + \beta_2 \alpha)]}} \quad (7)$$

where $\mathbf{a}$, $\boldsymbol{\alpha}$, and $\boldsymbol{\omega}$ represent peak linear acceleration, peak rotational acceleration, and peak rotational velocity, respectively. Bivariate models were also used to assess the relative contribution of linear and rotational information for injury risk prediction, allowing us to directly test our hypothesis that rotational acceleration is more predictive of concussion than linear acceleration. To do this, odds ratios and their 95% confidence intervals were compared between the linear and rotational term for each bivariate model. Predictor variables were standardized to have a mean of 0 and a standard deviation of 1, enabling comparisons on a common scale. Odds ratios were defined as the exponentiated logistic regression coefficients for a one-standard-deviation increase in each predictor. The Wald Test was used to assess significant differences between odds ratios for bivariate models including a linear and rotational term, with a Bonferroni-adjusted significance level of $p<0.017$. This approach allowed us to determine whether linear or rotational kinematics were more strongly associated with concussion outcome.

Additionally, we tested our hypothesis by comparing AUPRC values on the 20% held-out test set across the five kinematic models: $\mathbf{a}$, $\boldsymbol{\alpha}$, $\boldsymbol{\omega}$, $\mathbf{a} + \boldsymbol{\alpha}$ (Eq. 6), and $\mathbf{a} + \boldsymbol{\omega}$ (Eq. 7). AUPRC differences between models were tested using the pr.test function in the usefun R package, which performs a stratified bootstrap comparison of precision-recall curves based on the Davis and Goadrich formulation.[52] A significance level of $p<0.05$ (two-tailed) was used. AUPRC was chosen due to the highly imbalanced nature of the dataset, as precision-recall more accurately reflects the model's capability to classify the minority class. Finally, sensitivity was assessed on a second held-out test set of four youth concussions to assess whether decision thresholds derived on adult and young adult data can detect youth injuries.

### *Laboratory helmet tests*

To evaluate whether the proposed concussion risk metric can discriminate differences in helmet impact performance, American football helmets were subjected to controlled laboratory impact tests. Impact testing was conducted using a pneumatic linear impactor and an anthropomorphic test headform. The linear impactor had a ram mass of 15.5 kg and a curved, nylon end cap. The ram struck a size medium NOCSAE headform while it was equipped with a commercially available American football helmet (LS2; LIGHT Helmets; Carlsbad, CA). Following the widely recognized Virginia Tech STAR Helmet

Ratings,[23] four impact locations (Back, Front, Front Boss, Side) and three impact speeds (3.1 m/s, 4.9 m/s, 6.4 m/s) were selected for testing (Figure 4).

The helmet model we tested consisted of a composite shell, an expanded polypropylene liner, and 17 energy absorbing fit pads spread throughout its interior (Fig. 4a). Two styles of fit pads were tested in individual helmets that were otherwise identical. The helmet's standard set of fit pads contained D3O (D3O; London, United Kingdom), a non-Newtonian polymer whose stiffness increases with strain rate and which has been widely used in personal protective equipment applications. A modified set of fit pads contained SoftShox technology (SoftShox; Vienna, VA), liquid-based shock absorbers that dissipate impact energy through velocity-dependent fluid flow. This concept has been previously evaluated for American football applications through prototyping and finite element modeling.[53,54] Both styles of fit pads had an identical exterior casing, thus controlling for friction between the headform surface and inner helmet surface, and identical dimensions (diameter = 55 mm, thickness = 18 mm), thus controlling for helmet fit to the headform. Similar to the Virginia Tech STAR protocol, two replicate tests were conducted for each combination of impact location, impact speed, and helmet design, thus totaling 48 impact tests.

Triaxial linear accelerations and angular velocities were collected from the headform's center of gravity by a 6DX PRO sensor (Diversified Technical Systems; Seal Beach, CA) for each test. Linear accelerations and angular velocities were both filtered at 200 Hz. Peak linear accelerations and peak angular velocities were used to compute the bivariate risk metric $P_{inj}(\mathbf{a} + \boldsymbol{\omega})$.

Three independent two-way ANOVA's (one at each at impact velocity of 3.1 m/s, 4.9 m/s, and 6.4 m/s) were performed to assess the main effects of helmet technology and impact location, as well as their interaction, on concussive injury risk. When significant effects were identified, post-hoc pairwise comparisons were conducted using Šidák corrections to control the family-wise error rate.

## Results

### *Concussive impacts have larger kinematic and strain magnitudes*

We characterized the kinematics and finite-element simulated strains from a dataset that included 3,805 non-concussive impacts and 47 concussive impacts, collected across 203 male and female athletes in American football, Australian football, rugby union, gymnastics, ice hockey, MMA and lacrosse (Table 1). Concussive impacts occurred in 40 athletes, with three American football players and one Australian football player sustaining multiple concussions. Loss of consciousness occurred in five Australian football cases, three MMA cases, one collegiate football case, and one youth football case. We calculated peak linear acceleration ($\mathbf{a}$), peak rotational acceleration ($\boldsymbol{\alpha}$), peak rotational velocity ($\boldsymbol{\omega}$), HIP, split into linear and rotational components, and MPS95.

Figure 1a shows the Stanford MiG2.0, one of the iMG technologies used in this study, and its MEMS sensor components for measuring linear and rotational head impact forces. Figure 1b shows that concussive impacts exhibit significantly greater $\mathbf{a}$, $\boldsymbol{\alpha}$, $\boldsymbol{\omega}$, and MPS95 compared to non-concussive impacts (***$p<0.0001$), and that youth concussion magnitudes are much lower than concussion data from older athletes for $\mathbf{a}$ and $\boldsymbol{\alpha}$. Summary statistics of impact magnitudes are provided in Table 2.

Figure 1c shows that the vast majority of impacts fall within a regime where brain deformation is expected to be more sensitive to rotational velocity than to rotational acceleration. This pattern is shown relative to the white reference line, which has a slope equal to the inverse of the brain's natural period ($\Delta t_n$), approximately 40 ms when averaged across coronal, sagittal, and axial anatomical planes.[55] This period corresponds to a natural frequency of approximately 25 Hz and represents the duration of one full oscillation of the brain when subjected to dynamic loading. Impacts above this line are indicative of short-duration pulses less than $\Delta t_n$, where deformation is primarily influenced by rotational velocity. The predominance of impacts above this line suggests that rotational velocity plays a dominant role in brain deformation in this population.

In Figures 1d and 1e, we show differences in brain strain patterns for concussive and non-concussive impacts. Specifically, we show the MPS95 per element averaged across impact types. In concussive impacts, elevated strains are observed in the cortical areas including the upper gray matter (0.36 ± 0.21) and cortical white matter (0.30 ± 0.12), as well as the deep white matter regions such as the basal ganglia (0.28 ± 0.10) and corpus callosum (0.48 ± 0.14). Non-concussive impacts were significantly lower in MPS95, ranging from 0.04 to 0.09. (Supplementary Table 1). The distribution of regional strains is shown in Supplementary Figure 2.

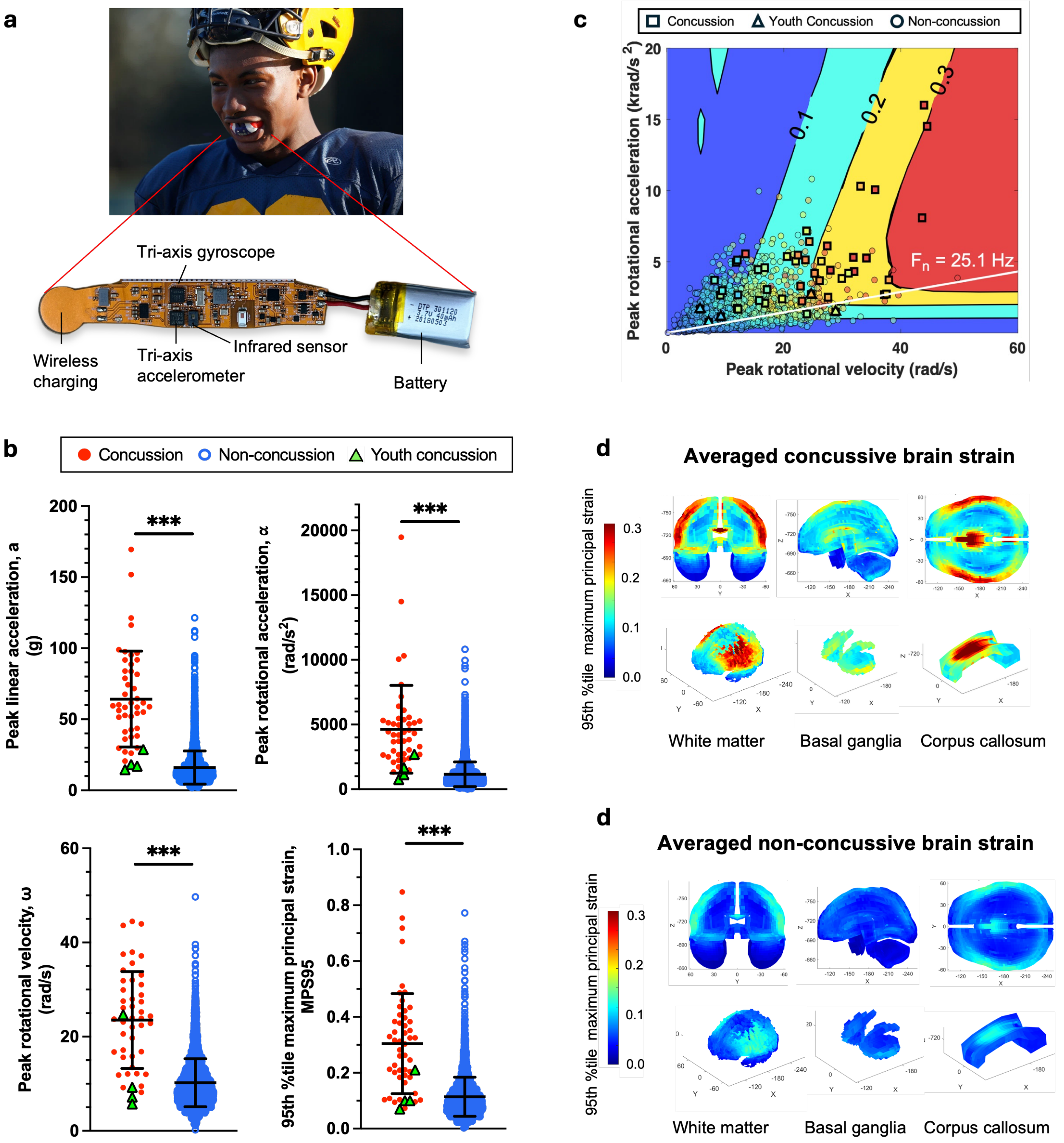

**Figure. 1. Differences in kinematics and brain strain between concussive and non-concussive impacts. a**, Stanford MiG2.0 and its sensor board components.[56] **b**, Distribution of impact magnitudes plotted on a logarithmic scale. Black bars represent the mean and one standard deviation of all impacts in each respective dataset. (***$p<0.0001$). **c**, Iso-strain contour plot showing the relationship between peak rotational velocity, peak rotational acceleration, and strain, as predicted by the DAMAGE model.[57] A white reference line is plotted with a slope defined as $\pi/\Delta t_n$, reflecting the brain's estimated natural frequency of 25.1 Hz. MPS95 patterns averaged across **d**, concussive and **e**, non-concussive impacts.

| | **Concussion** | | | | | | | | **Non-concussion*** | | | |
|---|---|---|---|---|---|---|---|---|---|---|---|---|
| | *Mean* | | *Median* | | *95th%* | | *SD* | | *Mean* | *Median* | *95th%* | *SD* |
| | *Non-Youth* | *Youth* | *Non-Youth* | *Youth* | *Non-Youth* | *Youth* | *Non-Youth* | *Youth* | | | | |
| **PLA (g)** | 68.3 | 19.7 | 60.2 | 17.6 | 120.7 | 27.2 | 32.2 | 6.2 | 16.0 | 12.6 | 39.0 | 11.7 |
| **PRA ($rad/s^2$)** | 4919 | 1560 | 4315 | 1399 | 10281 | 2536 | 3402 | 841 | 1156 | 899 | 2941 | 956 |
| **PRV (rad/s)** | 24.6 | 11.7 | 23.9 | 8.2 | 43.1 | 22.4 | 9.8 | 8.8 | 10.2 | 9.2 | 20.3 | 5.1 |
| **$HIP_{total}$ (kW)** | 10 | 0.78 | 8.4 | 0.79 | 31.7 | 1.1 | 8.9 | 0.33 | 1.0 | 0.6 | 3.7 | 1.7 |
| **$HIP_{lin}$(kW)** | 9.5 | 0.67 | 8.1 | 0.76 | 27.5 | 0.82 | 8.6 | 0.22 | 1.0 | 0.5 | 3.5 | 1.6 |
| **$HIP_{rot}$(kW)** | 1.5 | 0.26 | 1.1 | 0.09 | 4.7 | 0.7 | 1.8 | 0.37 | 0.1 | 0.1 | 0.4 | 0.2 |

**Non-concussive impacts do not include youth data.*

**Table 2.** Kinematics summary statistics for concussive and non-concussive impacts.

### *Linear acceleration is the most precise predictor variable for concussion*

Twelve kinematic and strain-based logistic regression models were developed to assess concussion classification performance and identify candidates for injury risk function development. Model deviance ($G^2$), Bayesian Information Criterion, LASSO-penalized logistic regression, and dominance analysis were used to screen all candidate predictors. All models had significantly better fit than the null model, but the five models with the best fit were $\mathbf{a} + \boldsymbol{\omega}$, $\mathbf{a} + \boldsymbol{\alpha}$, $\mathbf{a}$, $\boldsymbol{\alpha}$, and $\boldsymbol{\omega}$ based on the lowest deviance and Bayesian Information Criterion (Fig. 2a). Additionally, $\mathbf{a}$, $\boldsymbol{\omega}$, and $\boldsymbol{\alpha}$ were consistently the most informative and best-fitting fitting predictors across both LASSO and dominance analyses, based on having the largest $R^2$ and general dominance scores, respectively (Supplementary Tables 8-9). Therefore, $\mathbf{a} + \boldsymbol{\omega}$, $\mathbf{a} + \boldsymbol{\alpha}$, $\mathbf{a}$, $\boldsymbol{\alpha}$, and $\boldsymbol{\omega}$ were retained for formal statistical comparison as classifiers and for injury risk function development. This approach both preserved statistical power and enabled testing of our primary hypothesis that rotational acceleration is a more precise predictor of concussion than linear acceleration.

Table 3 summarizes the classification performances on the youth and held-out test sets. No concussions were detected on the youth set. On the held-out set, $\mathbf{a}$ achieved the highest AUPRC (0.65) and F1 score (0.50). Among univariate models, $\mathbf{a}$ was followed by $\boldsymbol{\alpha}$ (AUPRC=0.36; F1=0.31) and $\boldsymbol{\omega}$ (AUPRC=0.35; F1=0.30). Differences between $\mathbf{a}$ and $\boldsymbol{\alpha}$ are significant (Figs. 2b-c). Additionally, $\mathbf{a} + \boldsymbol{\alpha}$ (AUPRC=0.60; F1=0.55) offered no detectable improvement over the use of $\mathbf{a}$ alone. Among bivariate models, $\mathbf{a} + \boldsymbol{\omega}$ had the largest AUPRC (AUPRC=0.64, F1=0.40), and significantly outperformed $\boldsymbol{\alpha}$ and $\boldsymbol{\omega}$ alone (Fig. 2b). It is important to note that the AUPRC of a random classifier, defined as the ratio of positive cases to total cases, is only 0.01, indicating that kinematic metrics alone still provide substantial discriminatory power. Dominance analysis and LASSO further confirmed $\mathbf{a}$ as the most informative predictor variable, with the highest general dominance score ($R^2$=0.108) and LASSO coefficient (β=0.91), followed by **ω** ($R^2$=0.065; β=0.842) and **α** ($R^2$=0.061; β=0.380).

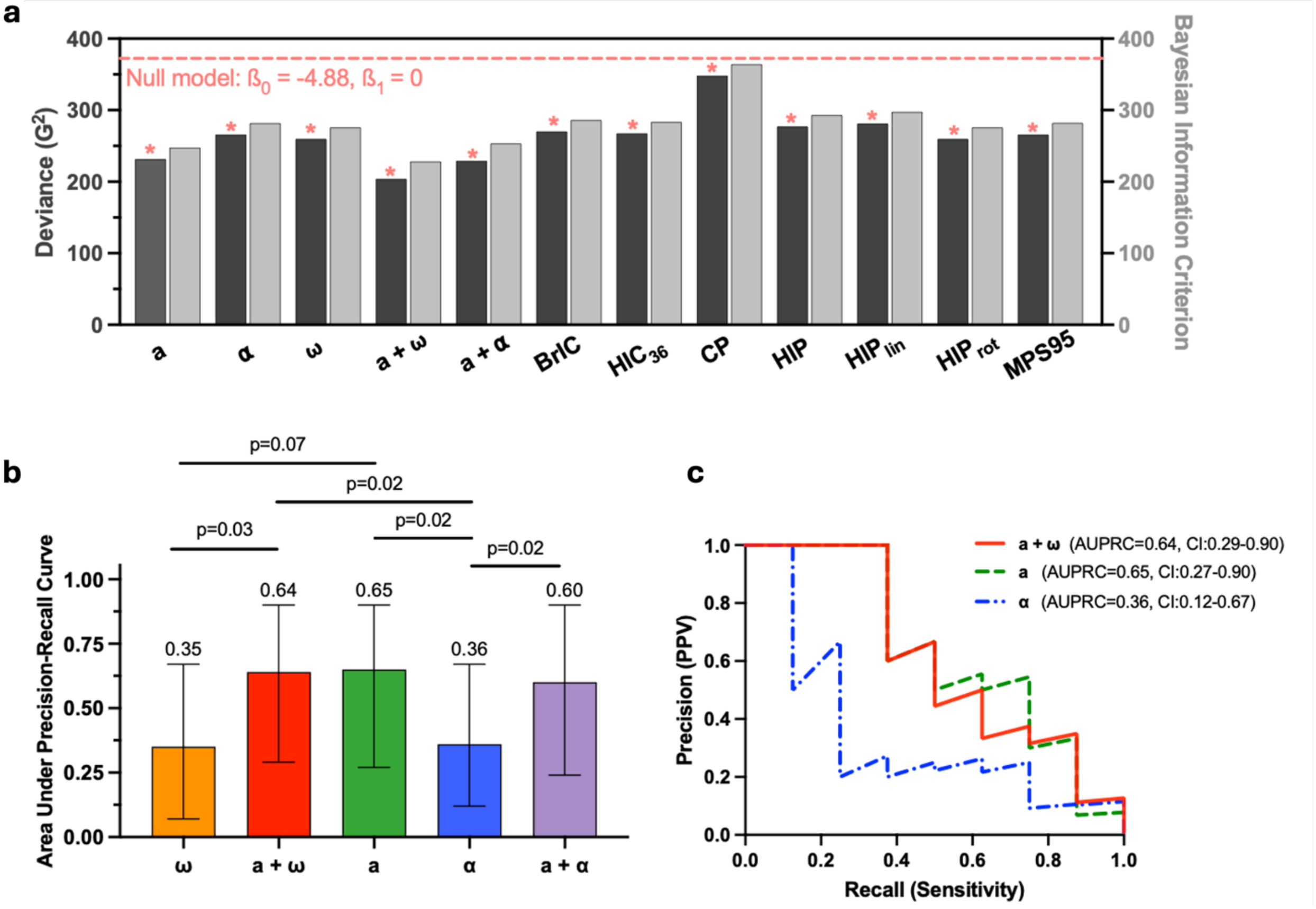

**Figure 2**. **Comparison of model fit and performance for concussion classifiers**. Model abbreviations are provided in Supplementary Table 2. **a,** In-sample goodness-of-fit for each model. A lower deviance ($G^2$) indicates better fit relative to the null model, and all models provided significantly better fit compared to the null model (*$p<0.005$). Bayesian Information Criterion assesses the tradeoff between model fit and complexity, with lower values indicating better performance while penalizing for model complexity. **b**, Area under the precision–recall curve for kinematic models, evaluated on the held-out test set. Black error bars represent 95% confidence intervals. **c**, Precision–recall curves for kinematic classifiers evaluated on the held-out test set.

Several additional metrics used in automotive and helmet safety assessments were evaluated, including MPS95, BrIC,[22] $HIC_{36}$,[49] HIP,[48] split into linear and rotational components, and Virginia Tech CP[50] (Supplementary Table 3). Although linear HIP outperformed rotational HIP based on AUPRC and F1, neither model achieved better fit than the five kinematic models described above. MPS95 and BrIC were among the lowest performers. As a result, these metrics were not used to construct injury risk functions or included in formal statistical comparisons.

| Model | Peak linear acceleration ($\mathbf{a}$) | Peak rotational acceleration ($\boldsymbol{\alpha}$) | Peak rotational velocity ($\boldsymbol{\omega}$) | Peak linear acceleration + peak rotational velocity ($\mathbf{a}$+ $\boldsymbol{\omega}$) | Peak linear acceleration + peak rotational acceleration ($\mathbf{a}$ + $\boldsymbol{\alpha}$) |
|---|---|---|---|---|---|
| **AUPRC [95% CI]**[1] | 0.65 [0.27, 0.90] | 0.36 [0.12, 0.67] | 0.35 [0.07, 0.67] | 0.64 [0.29, 0.90] | 0.60 [0.24, 0.90] |
| **F1 [95% CI]**[1] | 0.50 [0.18, 0.80] | 0.31 [0.13, 0.59] | 0.30 [0.13, 0.57] | 0.4 [0.22, 0.77] | 0.55 [0.25, 0.8] |
| **Precision**[1] | 0.80 | 0.33 | 0.33 | 1.0 | 0.5 |
| **Recall**[1] | 0.38 | 0.25 | 0.25 | 0.25 | 0.63 |
| **Recall (Youth)**[2] | 0 | 0 | 0 | 0 | 0 |
| **Decision Threshold**[1] | 0.2 (78.7 g) | 0.1 (5874.2 rad/s$^2$) | 0.1 (29 rad/s) | 0.3 | 0.1 |

*1: Held-out set, 2: Youth set*

**Table 3.** Performance metrics for logistic regression-based injury risk functions. The first test was conducted on the 20% held-out dataset, while recall (sensitivity) was calculated for the dataset containing 4 youth concussions.

### *Linear acceleration shows a significant association with concussion risk*

The five kinematic models identified by best fit were then used to develop injury risk functions. As described above, these included three univariate models ($\mathbf{a}$, $\boldsymbol{\omega}$, $\boldsymbol{\alpha}$) and two bivariate models ($\mathbf{a}$ + $\boldsymbol{\alpha}$ and $\mathbf{a}$ + $\boldsymbol{\omega}$). Model coefficients are provided in Supplementary Table 4, and the resulting risk functions are shown in Figure 3. The 50% injury risk thresholds were determined to be 100 g for $\mathbf{a}$, 8.3 krad/s$^2$ for $\boldsymbol{\alpha}$ and 40 rad/s for $\boldsymbol{\omega}$ (Fig. 3a–c).

Next, bivariate models allowed us to assess the relative importance of linear versus rotational terms. In the $\mathbf{a}$ + $\boldsymbol{\alpha}$ model, the 50% risk contour displayed a steep slope, suggesting greater sensitivity to $\mathbf{a}$. Consistent with this, odds ratios (OR) indicated $\mathbf{a}$ was a significant independent predictor of concussion (OR=2.3; 95% CI: 1.5-3.6), while $\boldsymbol{\alpha}$ was not (OR=1.3; 95% CI: 0.81-1.9), as its confidence interval included 1. The larger odds ratio for $\mathbf{a}$ indicates it is a greater risk factor for concussion than $\boldsymbol{\alpha}$. However, the Wald test comparing the two ORs was not statistically significant (Fig. 3f). In the $\mathbf{a}$ + $\boldsymbol{\omega}$ model, risk increased comparably with both predictors, as reflected by their similar odds ratios ($OR_{\mathbf{a}}$=2.1; $OR_{\boldsymbol{\omega}}$=2.2) with no significant differences (Fig. 3f).

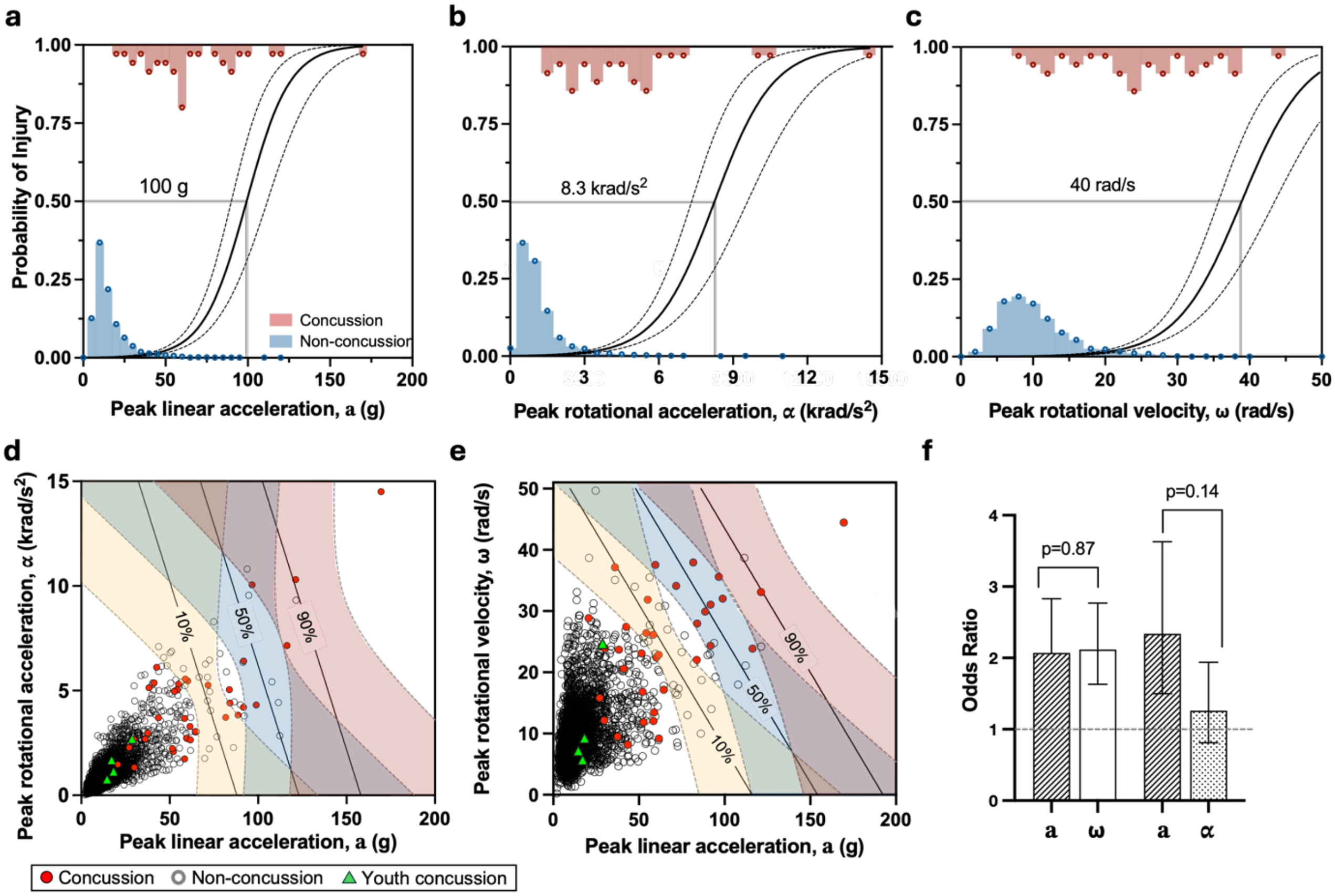

**Figure 3. Injury risk functions developed on the 80% training data**. Univariate logistic regression models based on **a**, peak linear acceleration, **b**, peak rotational acceleration, and **c**, peak rotational velocity. Solid black lines indicate the fitted logistic regression curves and dashed lines show the 95% confidence intervals. Injury risk functions for **d**, **a + α** and **e, a + ω**. The black contours and shaded regions indicate different risk levels and their 95% confidence intervals. Youth concussions are overlaid for comparison and did not contribute to the training data. **f**, Odds ratios plotted for each predictor in bivariate kinematic models. In the **a + α** model, the confidence interval for **α** crosses 1, indicating it is not a significant independent predictor of concussion.

### *Liquid helmet technology significantly reduces concussion risk*

Helmets equipped with novel liquid-filled fit pads were evaluated against a control helmet equipped with stock fit pads using the impact speeds and locations defined in the Virginia Tech Varsity Football Helmet STAR protocol (Fig. 4a-b). Concussion risk was quantified using the best-performing bivariate injury metric $P_{inj}(a + \omega)$. Across impact velocities, the liquid helmet exhibited consistently lower injury risk compared to the control helmet (Fig. 4c-e). At 3.1 m/s, mean injury risk averaged across impact locations was 1.6% for the Control helmet and 0.8% for the Liquid helmet (Fig. 4c), with statistically significant differences between helmets based on post-hoc comparisons with Šidák correction.

At 4.9 m/s, mean injury risk decreased from 26.7% to 13.4% with the liquid pad, with significant reductions at the back, front, and side locations (Fig. 4d). At the highest impact velocity of 6.4 m/s, the liquid helmet reduced mean injury risk from 73.3% to 55.7%, representing the largest observed reduction in absolute risk, with significant reductions at the back, front, and side locations (Fig. 4e). Additionally, the bottom row of Figure 5 shows that the reductions in injury risk are primarily driven by attenuation of peak linear acceleration. Peak kinematics and associated risk predictions for individual impacts are provided in Supplementary Table 8.

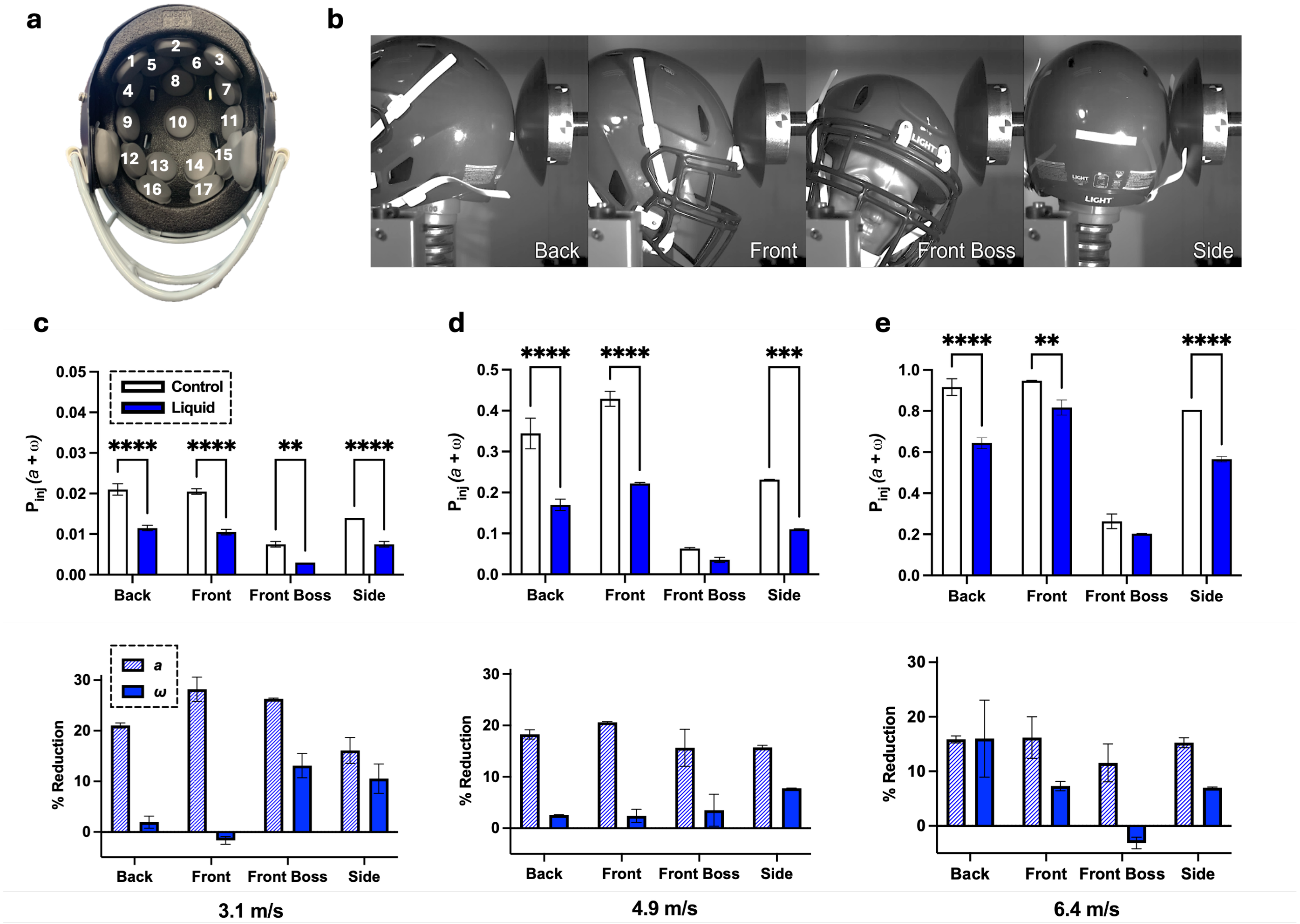

**Figure 4. Comparison of helmet technologies for mitigating concussion risk**. **a,** Fit pad locations adapted with liquid shock absorbers. **b,** Impact locations tested in the laboratory. **c-e,** Injury risk quantified as $P_{inj}(a + \omega)$ is shown for helmets equipped with standard fit pads (Control) and liquid fit pads (Liquid), as well as percent reductions (Liquid – Control) in $a$ and $\omega$ across impacts with velocities of **c**, 3.1 m/s, **d**, 4.9 m/s, and **e**, 6.4 m/s. Bars represent the mean ± standard deviation. Asterisks denote with statistically significant reductions observed at each location (**p < 0.01, ***p < 0.001, ****p < 0.0001).

## Discussion

The rotational acceleration hypothesis for concussion has guided brain injury prevention research, shaping safety standards and the development of protective equipment. Yet, this hypothesis had not been tested using direct measurements of rotational acceleration during concussion in humans. We addressed this gap by collecting a dataset of directly measured rotational accelerations during human concussions using instrumented mouthguards across multiple sports and age groups. Surprisingly, we found rotational acceleration was not a significant independent predictor of concussion risk, and was significantly outperformed in classification ability by linear acceleration. We propose two new injury risk functions incorporating peak linear acceleration. For the univariate risk function, we found a 50% concussion risk threshold of 100 g, when filtered at 200 Hz. The bivariate risk function incorporates peak linear acceleration and peak rotational velocity, as both variables were found to be significant independent predictors of concussion. Applying this risk function to football helmet impact tests, we demonstrate that concussion risk can still be meaningfully reduced using liquid-padding technology. Together, these findings indicate that linear acceleration is a primary risk factor for concussion and that it can be a target metric to improve helmet safety.

Given the utility of single-impact thresholds in helmet and vehicle safety evaluations, we restricted our analysis to the kinematics of the impact linked to each diagnosed concussion and evaluated injury classification and risk using single-event measures. Although repeated head impacts are known to contribute to long-term neurodegenerative disease,[58,59] the extent to which prior exposure modifies instantaneous risk remains debated, and both single-impact and cumulative-impact modeling approaches are ongoing areas of investigation. Prior studies conducted by the National Collegiate Athletic Association and Department of Defense Grand Alliance[60] have shown that injury may be preceded by either a high-magnitude single impact or clusters of lower-magnitude impacts, suggesting that multiple mechanisms may exist for acute concussion.[61] However, safety evaluations typically define pass or fail criteria that are based on individual loading events rather than accumulated exposure, and protective equipment is assessed relative to these thresholds.[50,62–65] As a result, our focus on single-event kinematics is aligned with prevailing safety evaluation practices, and remains an appropriate and widely used framework for estimating acute concussion risk.

Using this single-event modeling approach, we found peak linear acceleration to be a significantly more precise predictor of concussion than peak rotational acceleration. Peak linear acceleration was also the largest risk factor for concussion and the only significant independent contributor in the model that also contained peak rotational acceleration ($\mathbf{a} + \boldsymbol{\alpha}$; Fig. 3d). Peak rotational velocity was also found to be a significant independent risk factor for concussion when evaluated alongside linear acceleration, and the combined model ($\mathbf{a} + \boldsymbol{\omega}$) achieved the second largest AUPRC. Dominance analysis and LASSO-penalized logistic regression further supported the predictive utility of peak linear acceleration as it contributed the most unique and stable predictive signal across both methods, followed by peak rotational velocity. A similar finding was observed in elite men's rugby union using iMGs, where peak linear acceleration was significantly associated with player removal for head injury assessment due to suspected concussion, while peak rotational acceleration was not.[66] However, conclusions from the rugby study are limited by the use of receiver operating characteristic curves for evaluating predictive performance, which does not assess positive predictive value (i.e., precision). In datasets with a low concussion incidence such as the rugby study (58 removals, 5,400 non-removals), the area under the receiver operating characteristic curve is less informative because it gives equal weight to both classes and can be dominated by correct predictions of the much larger negative class (i.e. impacts without removal from play). In contrast, we used precision−recall, which includes positive predictive value and sensitivity, and offers a more balanced evaluation of the classifier’s performance on rare outcomes like concussion.

Likewise, a previous study using HITS football data reported better predictive performance for peak linear acceleration compared to peak rotational acceleration on a dataset of 244 concussions and 62,767 non-concussions.[50] However, to address underreporting, the concussion dataset in the HITS study included non-concussive impacts that were relabeled as concussions based on having large kinematic magnitudes. Additionally, there were no significant differences between peak linear acceleration and peak rotational acceleration in a separate test dataset on professional football impact reconstructions.[50] The HITS study similarly quantified predictive performance using receiver operating characteristic curves, where the area under the curve is driven by the correct classification of the majority negative class of non-concussions. Taken together, these factors complicate the direct interpretation of rotational acceleration as a predictor of clinically diagnosed concussions in the HITs dataset.

An intriguing outcome of the current study is that strain-based and composite injury metrics underperformed simple kinematic measures. MPS95 and BrIC which are commonly used in research and regulatory settings. MPS95 is the target output of the Diffuse Axonal Multi-Axis General Evaluation (DAMAGE) model, a second-order system for estimating maximum brain strain used in the European New Car Assessment Program safety rating.[65] The National Football League also uses DAMAGE alongside the $HIC_{36}$, which is based on linear acceleration, to calculate Helmet Performance Score for helmet safety rankings. Although the Helmet Performance Score has been shown to correlate with on-field concussion incidence,[67] we found that MPS95 had low classification performance compared to kinematic metrics, including $HIC_{36}$ (Supplementary Table 3). Given that $HIC_{36}$ performed well and similarly to peak linear acceleration in our dataset, the observed correlation of Helmet Performance Score with concussion incidence may be driven by its linear acceleration component, $HIC_{36}$, rather than DAMAGE. Strain prediction models are also dependent on rotational acceleration and rotational velocity inputs, in agreement with Holbourn's hypothesis. We acknowledge that peak rotational acceleration measurements may be subject to higher error rates due to the differentiation of rotational velocity. However, Figure 1b shows that the majority of impacts fall within the loading regime where brain deformation is more sensitive to rotational velocity than rotational acceleration, so it is less plausible that noise in the rotational acceleration input diminished the performance of MPS95. Alternatively, the relatively low performance of MPS95 may reflect limitations in brain model fidelity, as strain-based bicycle helmet rankings have been shown to differ widely across brain models,[68] [68]suggesting our findings may be model-dependent. Regional strain metrics have also been shown to improve classification,[69] though we only evaluated global MPS95 here. BrIC, a rotational velocity-based metric proposed by the National Highway Traffic Safety Association for automotive crash testing, has similarly informed helmet safety evaluations[64], however BrIC had the lowest predictive performance in our analysis (Supplementary Table 3). Prior work has shown that the critical rotational velocity thresholds used in BrIC can be exceeded during voluntary head motions without producing injurious brain strain.[70] This may have led to false positives during classification, which could explain BrIC's relatively lower performance here.

Interestingly, peak rotational velocity was a significant independent predictor of concussion when evaluated alongside linear acceleration. This may reflect that most impacts in our dataset were short-duration and fell within the rotational velocity-sensitive regime for brain deformation, rather than the rotational acceleration-sensitive regime (Fig. 1b). However, this interpretation is likely duration-dependent. Under longer-duration loading, high rotational velocities may occur without sufficient rotational acceleration to cause injury, as demonstrated during voluntary head rotations and figure skating.[70,71] Therefore, rotational velocity-based metrics may overpredict injury risk for longer-duration head motions that differ from those represented in our dataset, such as whiplash-like loading.

Given the predictive utility of linear acceleration, we speculate one biomechanical mechanism by which linear acceleration could directly contribute to concussion. Due to the slight density difference between the brain and cerebrospinal fluid (CSF), rapid linear acceleration of the skull may cause differential motion of the brain as it inertially lags behind. When defining the acceleration input as a half-sine pulse, we can estimate the relative displacement of the brain within the skull as

$$x_{rel}(T) = \left(\frac{\rho_{brain} - \rho_{CSF}}{\rho_{brain}}\right) \cdot \frac{AT^2}{\pi}$$

where $\rho$ , A, and T represent density (kg/m$^3$) acceleration (m/s$^2$) and time (ms), respectively. Assuming a 100 g acceleration pulse over 10 ms, and densities of 1000 kg/m$^3$ and 1040 kg/m$^3$ for $\rho_{brain}$ and $\rho_{CSF}$, respectively, we estimate the brain would the brain would translate approximately 1.2 mm relative to the skull. At these higher acceleration levels, differential brain motion could compress the subarachnoid space and rapidly displace CSF through available outflow pathways such as the foramen magnum, thereby limiting its shock-absorbing capacity and potentially contributing to injury of surrounding structures. We speculate that this displacement might lead to injury due to contact forces between the cortex and the skull, or perhaps at the depths of sulci[59,72,73] as the gyri flex from motion/contact. Alternatively, negative pressure at the contrecoup site could induce cavitation, a phenomenon in which vapor bubbles form and collapse within the CSF, blood, or parenchyma due to rapid pressure changes. Because pressure has shown to correlate strongly with linear acceleration,[13] this mechanism may be relevant in high-acceleration impacts. Cavitation is also a leading hypothesis for TBI from military blast exposure,[74–76] which could support linear acceleration as a unifying theory for blast-related brain injuries and concussion. However, this warrants future investigation as there is limited evidence supporting the occurrence of cavitation *in vivo*. Future studies should investigate brain displacement, pressure gradients, and fluid flow during pure linear acceleration using fluid-structure interaction models.[77,78] This would also address limitations of current finite element brain models, as many treat CSF as a solid and may not capture fluid dynamics. This limitation may also explain why MPS95 had relatively lower predictive ability here. To add, we do not dispute that rotational motion of the head can cause diffuse brain injury, especially given the prevalence of animal studies[17,18,20,79] that have shown correlation between angular acceleration and axonal pathology. However, our data suggests that alternative mechanisms could be at play.

We derived a 50% concussion risk threshold of approximately 100 g (95% CI: 89g−110g) for peak linear acceleration (Fig. 3a). This threshold falls in between prior estimates based on on-field studies. Pellman et al. reported a lower 50% risk level of 85 g based on laboratory reconstructions of 31 impacts (25 concussions, 6 non-concussions) from the National Football League.[63] This could be attributed to the artificially large concussion incidence from selection bias, potentially overestimating the model's ability to distinguish between concussion and non-concussion cases and skewing the risk curve toward lower thresholds. The lower threshold may also reflect methodological limitations of laboratory reconstructions, such as idealized impact conditions and reliance on video resolution for kinematic accuracy. Similarly, McIntosh and colleagues reported a lower 50% risk threshold of 65 g in Australian football using computational multibody reconstructions of 40 impacts (27 concussions, 13 non-concussions),[80] where overrepresentation of concussive cases likely contributed to selection bias and a risk function biased toward lower accelerations. By contrast, HITS data estimated 50% concussion risk at 192 g[23] which is much higher than our observed threshold. This discrepancy may be due to noise introduced by relative motion between the helmet and skull, or the use of a higher filter cutoff frequency (CFC1000), which could have preserved high-frequency signal or noise components.[24] Additionally, high acceleration false positive impacts due to lack of visual-verification in the non-concussive data could have been present, potentially causing the logistic regression model to learn a decision boundary skewed toward higher magnitudes and shifting estimated risk thresholds upward.

Many safety testing standards permit peak linear accelerations that are much higher than the 50% risk threshold derived in the current study, reflecting historical prioritization for preventing catastrophic head injuries rather than concussion.[81] A recent exception is the Fédération Équestre Internationale proposal to include a 150 g threshold specifically for equestrian helmet concussion testing; however, our model indicates this corresponds to a >99% probability of injury.[82] Based on Virginia Tech laboratory testing of 45 different equestrian helmets, compliance with this proposed standard would result in only 4 helmets (9%) below this 150g threshold.[83] If our model is correct, even these few higher performing helmets would not meaningfully reduce concussion risk. A helmet threshold of 100 g (the 50% risk level with a 200 Hz filter), however, would likely improve safety but requires innovation, which may indeed be possible depending on the energies involved in exposure.

The exposure in football has been better characterized than any other helmet type, so we evaluated the ability of a novel, liquid-filled technology to reduce concussion risk during laboratory football helmet impacts. We characterized risk using the $P_{inj}(a + \omega)$ bivariate function—given its superior high discriminatory power—and because rotational kinematics were more realistically modeled in the laboratory with the inclusion of the neck and sliding torso mass. The helmet equipped with liquid-filled fit pads exhibited significantly lower predicted concussion risk than the control helmet, primarily driven by attenuation of peak linear acceleration (Figure 4). Across impact velocities, the liquid helmet most consistently reduced risk at the back location, followed by the side, front, and front boss location, which had the fewest significant reductions. Notably, the back and side locations have been identified as common impact sites for concussion,[32,84] and are frequent during helmet-to-ground impacts, which account for 18% of concussions in the National Football League.[85] Additionally, the improved performance of the liquid helmet is likely due in part to greater compression of the liquid shock absorbers, prototypes of which have been previously shown to enable more stroke to be utilized as the head decelerates.[53] As the shock absorbers compress, fluid is forced through an orifice and pressure builds in proportion to the orifice area and flow rate, the goal of which is to create a regulated resistive force that remains steady throughout the stroke and reduces sharp acceleration spikes.[54,86,87] The shock-absorbing effect of CSF may protect the brain in a similar manner though this warrants future investigation. CSF may also allow frictionless gliding between the brain and skull during head rotation.[88,89] Reflecting this biological motivation, reducing friction at the head–helmet interface to minimize torque and rotational kinematics has become an important focus of several contemporary concussion mitigation technologies, including slip-liner and deformable liner systems such as MIPS, WaveCel, and SPIN.[90] While these approaches represent progress in addressing rotational components of head motion, our finding that linear acceleration is the most significant predictor of concussion suggests that minimizing force transmission to the head may be at least as important as minimizing torque for risk mitigation.

The current dataset included directly measured head kinematics from concussive impacts across children and adult athletes of both sexes. While we did not conduct sex-specific analyses, this remains an important area for future research, particularly given existing evidence that females may be more susceptible to concussion and axonal injury than males.[20,91–93] Our dataset only included three female concussions from athletes in lacrosse, rugby, and women's artistic gymnastics. Additionally, youth concussions had much lower magnitudes compared to those from older athletes (Table 2). The average peak linear acceleration was 19.7 g for youth and 68.3 g for non-youth concussions, while the average peak rotational acceleration was 1,560 rad/s$^2$ for youth and 4,919 rad/s$^2$ for non-youth. During testing, the kinematic classifiers could not detect any youth concussions (Table 3), suggesting that much lower thresholds may be needed to accurately assess injury risk in youth athletes and develop effective helmets. Future work is needed to characterize kinematic and strain-based injury thresholds in youth populations using iMGs across a broad range of sports.

This study has several limitations that warrant consideration. First, head impact data were collected using four different iMGs, each validated independently with device-specific filtering and cutoff frequencies. To ensure consistency across devices, kinematic traces were low-pass filtered at 200 Hz. This decision was guided by the fact that roughly half of the dataset came from Prevent iMG data, which were only available with 200-Hz filtering. Further analysis also revealed that adjusting for iMG device and sport in the logistic regression had negligible effects on risk predictions (Supplementary Information). Second, there is some variability in the selection of concussive impacts due to the inclusion of MMA, where concussive impacts were selected based on having the largest peak rotational acceleration prior to diagnosis. Given that our primary finding is that peak linear acceleration is the more precise classifier, this selection approach is unlikely to have biased the results in favor of linear acceleration. Third, there is also a large difference, naturally, between the number of concussive versus non-concussive impacts in this study, which makes classification more challenging. Even so, the 47 concussions reported here represent, to our knowledge, the largest and most diverse iMG-collected concussion dataset assembled to date. Fourth, the predictive accuracy of our thresholds and predicted reduction in concussion risk would also need to be determined in a prospective study. Finally, our injury risk functions were derived from sports impacts, which largely occupied a short-duration loading regime. As a result, these risk functions should not be extrapolated to longer-duration head motions, such as whiplash-like loading or non-impact head motions, without additional validation. This distinction is also relevant when comparing our findings with animal studies of rotational injury. Animal models have different head and brain geometry and material properties from humans, which has motivated frequency- rather than kinematics-based scaling approaches.[94] The relative importance of rotational acceleration, rotational velocity, and linear acceleration may therefore depend on species-specific brain dynamics as well as loading duration and frequency content.[55,94] For example, a porcine model of traumatic axonal injury showed that peak rotational acceleration was more strongly correlated with axonal injury volume than peak rotational velocity across multiple loading directions.[95] Future work should define the broader magnitude-duration conditions under which longer-duration head motions cause human concussion.

To our knowledge, this is the first study to evaluate the rotational acceleration hypothesis using *in vivo* head kinematics from a dataset reflecting measured concussion incidence. While rotational kinematics have guided concussion prevention because of the brain's susceptibility to shear deformation, our findings suggest that rotational acceleration alone may be insufficient for concussion risk assessment. Across a multi-center dataset of video-verified impacts collected from male and female athletes, peak linear acceleration was a significantly more precise classifier of concussion than rotational acceleration and remained a significant independent risk factor. At the same time, peak rotational velocity remained a significant independent risk factor when considered jointly with linear acceleration, supporting a multivariate risk model that incorporates still both linear and rotational head motion. We also identified a 50% concussion risk level of approximately 100 g for peak linear acceleration, lower than values historically considered in regulatory test standards. Finally, applying the resulting injury risk function to laboratory football helmet evaluations showed that predicted concussion risk could be meaningfully reduced using liquid-based shock absorption technology. Together, these findings support concussion safety evaluations that explicitly account for linear acceleration to better guide helmet design and injury risk reduction.

## Declarations

*Ethics approval*

This study was approved by the Stanford University Institutional Review Board (Protocol 45932), the Monash University Human Research Ethics Committee (Project IDs: 27684, 36297, 36556) and Deakin University (Project ID: 2022-337). Clinical trial number not applicable.

*Consent*

All authors have agreed with the content of this manuscript and have given explicit consent to submit the manuscript to Annals of Biomedical Engineering.

*Data, Material, and/or Code availability*

The datasets analyzed during the current study are available from the corresponding author on reasonable request.

*Competing interests*

D.B. Camarillo's contribution to this publication regarding the mouthguard work was Stanford research, but the helmet work was a result of his being a paid investigator by SoftShox via an NIH Phase 1 SBIR collaboration with Stanford (R43NS119134). Additionally, N.J. Cecchi and D.B. Camarillo are inventors of and have financial interests in the liquid shock absorbing technology which is covered by pending patent applications from both Stanford University and SoftShox. All other authors declare no competing interests.

*Funding*

Funding for this study was provided by the Taube Stanford Children's Concussion Initiative, Pac-12 Conference's Student-Athlete Health and Well-Being Initiative, the National Institutes of Health (R24NS098518; R43NS119134; R44NS119134), the Stanford Maternal and Child Health Research Institute, the Office of Naval Research, and the National Science Foundation Graduate Research Fellowship Program.

*Authors' contributions*

D.B.C. and J.A.T. contributed to the conceptualization and initiation of the study. D.B.C., M.M.Z., S.J.M., and J.A.T. contributed to the methodological approach. M.M.Z, G.A.G. and D.B.C. initiated field data collection at Stanford. J.A.T. led the study design and performed all data analyses. S.J.M. oversaw the collection of Australian football data. J.W.H., W.T.O., and S.S.H.R. collected and processed the Australian football data. J.E.U., J.D.S., and N.S.P. contributed Wake Forest concussion data. N.S.P. processed the Wake Forest concussion kinematic data. N.J.C. contributed the laboratory helmet testing data. J.A.T. wrote the manuscript. D.B.C. supervised the analysis and manuscript writing. All authors provided comments on the manuscript.

*Acknowledgements*

The authors want to thank Drs. Joel Stitzel, Jill Urban, and Nicholas Stewart Pritchard for contribution of the youth concussion data. We also thank Drs. Steven Rowson, Vahidullah Tac, and Xianghao Zhan for providing help with the methodological approach. We thank Dr. Fady Abayazid for providing feedback on the manuscript. We also thank Drs. Matthew Campbell,

Lyndia Wu, Fidel Hernandez, Kaveh Laksari, Michael Fanton, Abigail Swenson, Ty Holcomb, and Madison Marks for data collection efforts. We thank Lauren Evans, Joel Ernest, William Zhou, Che Bonini, Mary-Anne Camenzuli, Luke Boykett, Jason Turner, Av Kumar, and Paul Kennedy for their Australian football data collection efforts. We thank Tie Liang for help with the statistical approach. We thank Abigail Huddleston, Angel Gunaman, and Praveen Sundar for assisting with video review. We thank Kyle Casazza and Quinn Gaughan for their efforts toward laboratory helmet impact testing. J.A.T. acknowledges support from the Stanford Graduate Fellowship, National Science Foundation Graduate Research Fellowship, Enhancing Diversity in Graduate Education Fellowship, and Stanford's Department of Bioengineering.

# Supplementary Data

**Supplementary Data Table 1.** Head impact datasets used in this study.

| Authors | Sport | Level(s) | Gender(s) | Measurement Device | No. Concussions | No. Non-concussions | Diagnostic criteria |
|---|---|---|---|---|---|---|---|
| Current study | Lacrosse | Collegiate | Female | Stanford MiG2.0 | 1 | 102 | Immediate diagnosis |
| Current study | Rugby | Collegiate | Female/ Male | Prevent iMG | 3 | 469 | Immediate diagnosis/ player identification of impact |
| Domel et al.[1] | American football | Collegiate | Male | Stanford MiG2.0 | 0 | 359 | |
| Hernandez et al.[2] | American football | Collegiate | Male | In-Mouth (ADXL377/ L3G4200D) | 1 | 0 | Immediate diagnosis |
| Bartsch et al.[3] | American football | High-school, collegiate | Male | Prevent iMG | 13 | 1782 | Visual signs of concussion |
| Tiernan et al.[4] | Mixed martial arts | Professional, semi-professional, amateur | Male | Stanford MiG 2.0 | 5 | 304 | Largest acceleration impact immediately prior to diagnosis |
| Evans et al.[5] and ongoing study | Australian football | Amateur | Male | HITIQ | 19 | 776 | Immediate diagnosis |
| NIH Grant K25HD101686 | American football | Youth | Male | WF | 3 | 0 | Immediate diagnosis |
| Pritchard et al.[6] | Gymnastics | Advanced | Female | WF | 1 | 0 | Immediate diagnosis |
| Swenson et al.[7] | Ice hockey | Youth | Male | WF | 1 | 0 | Immediate diagnosis |

**Supplementary Data Table 2.** Kinematics summary statistics for concussive and non-concussive impacts.

| | Concussion | | | | | | | | Non-concussion | | | |
|---|---|---|---|---|---|---|---|---|---|---|---|---|
| | *Mean* | | *Median* | | *95th%* | | *SD* | | *Mean* | *Median* | *95th%* | *SD* |
| | *Non-Youth* | *Youth* | *Non-Youth* | *Youth* | *Non-Youth* | *Youth* | *Non-Youth* | *Youth* | *Non-Youth* | *Non-Youth* | *Non-Youth* | *Non-Youth* |
| **PLA (g)** | 68.3 | 19.7 | 60.2 | 17.6 | 120.7 | 27.2 | 32.2 | 6.2 | 16.0 | 12.6 | 39.0 | 11.7 |
| **PRA (rad/s$^2$)** | 4919 | 1560 | 4315 | 1399 | 10281 | 2536 | 3402 | 841 | 1156 | 899 | 2941 | 956 |
| **PRV (rad/s)** | 24.6 | 11.7 | 23.9 | 8.2 | 43.1 | 22.4 | 9.8 | 8.8 | 10.2 | 9.2 | 20.3 | 5.1 |
| **$HIP_{total}$ (kW)** | 10 | 0.78 | 8.4 | 0.79 | 31.7 | 1.1 | 8.9 | 0.33 | 1.0 | 0.6 | 3.7 | 1.7 |
| **$HIP_{lin}$(kW)** | 9.5 | 0.67 | 8.1 | 0.76 | 27.5 | 0.82 | 8.6 | 0.22 | 1.0 | 0.5 | 3.5 | 1.6 |
| **$HIP_{rot}$(kW)** | 1.5 | 0.26 | 1.1 | 0.09 | 4.7 | 0.7 | 1.8 | 0.37 | 0.1 | 0.1 | 0.4 | 0.2 |

**Supplementary Data Table 3.** Regional strain summary statistics for concussive and non-concussive impacts.

| | Concussion | | | | Non-concussion | | | |
|---|---|---|---|---|---|---|---|---|
| | *Mean* | *Median* | *95th%* | *SD* | *Mean* | *Median* | *95th%* | *SD* |
| **Global** | 0.32 | 0.32 | 0.70 | 0.18 | 0.12 | 0.11 | 0.26 | 0.07 |
| **White matter** | 0.30 | 0.30 | 0.65 | 0.17 | 0.12 | 0.10 | 0.26 | 0.07 |
| **Upper gray matter** | 0.36 | 0.35 | 0.80 | 0.21 | 0.14 | 0.12 | 0.29 | 0.08 |
| **Lower gray matter** | 0.20 | 0.20 | 0.39 | 0.11 | 0.08 | 0.07 | 0.16 | 0.04 |
| **Thalamus** | 0.27 | 0.24 | 0.56 | 0.20 | 0.08 | 0.07 | 0.18 | 0.06 |
| **Brainstem** | 0.23 | 0.21 | 0.44 | 0.16 | 0.06 | 0.15 | 0.04 | 0.08 |
| **Corpus callosum** | 0.48 | 0.36 | 1.28 | 0.41 | 0.11 | 0.08 | 0.27 | 0.10 |
| **Basal ganglia** | 0.28 | 0.26 | 0.54 | 0.16 | 0.10 | 0.09 | 0.22 | 0.06 |
| **Cerebellum** | 0.13 | 0.13 | 0.21 | 0.07 | 0.04 | 0.04 | 0.09 | 0.02 |

**Supplementary Data Table 4.** Kinematic and strain-based model abbreviations referenced in this study.

| Metric | Abbreviation |
|---|---|
| Peak linear acceleration | **a** |
| Peak rotational velocity | **ω** |
| Peak rotational acceleration | **α** |
| 95th percentile maximum principal strain | MPS95 (%) |
| Head injury criterion[8] | $HIC_{36}$ |
| Brain injury criterion[9] | BrIC |
| Combined probability[10] | CP |
| Head impact power[11] | HIP (kW) |
| Linear head impact power | $HIP_{lin}$ (kW) |
| Rotational head impact power | $HIP_{rot}$ (kW) |

**Supplementary Data Table 5**. Performance metrics for additional kinematic- and strain-based classifiers.

| Metric | AUPRC | F1 | Precision | Recall |
|---|---|---|---|---|
| **MPS95** | 0.28 | 0.29 | 0.33 | 0.25 |
| **HIP** | 0.51 | 0.50 | 0.75 | 0.38 |
| **$HIP_{lin}$** | 0.51 | 0.50 | 0.75 | 0.38 |
| **$HIP_{rot}$** | 0.39 | 0.35 | 0.50 | 0.25 |
| **$HIC_{36}$** | 0.60 | 0.36 | 0.66 | 0.25 |
| **BrIC** | 0.18 | 0.25 | 0.22 | 0.25 |
| **CP** | 0.45 | 0.36 | 0.67 | 0.25 |

**Supplementary Data Table 6.** Coefficients for injury risk functions.

| Model | Intercept | B1 | B2 |
|---|---|---|---|
| **Peak linear acceleration** | -8.363 | 0.0904 | - |
| **Peak rotational acceleration** | -6.966 | 0.00084 | - |
| **Peak rotational velocity** | -8.785 | 0.2353 | - |
| **Peak linear acceleration + peak rotational velocity** | -9.946 | 0.0741 | $1.38 \times 10^{-1}$ |
| **Peak linear acceleration + peak rotational acceleration** | -8.327 | 0.0949 | $-7.89 \times 10^{-5}$ |

**Supplementary Data Table 7.** Specifications for instrumented mouthguards used in this study.

| | In-mouth (ADXL377/ L3G4200D) | Stanford MiG2.0 | Prevent iMG | HITIQ Nexus | Wake Forest Mouthpiece |
|---|---|---|---|---|---|
| **Sampling rate (Accel)** | 1024 Hz | 1000 Hz | 3200 Hz | 3200 Hz | 350[a] , 4684[b], & 6400[c] Hz |
| **Sampling rate (Gyro)** | 1024 Hz | 8000 Hz | 3200 Hz | 800 Hz | 350[a], 1565 [b], & 6400[c] Hz |
| **Measurement range (Accel)** | ± 200 g (1 tri-axis) | ± 400 g (1 tri-axis) | ± 200 g (1 tri-axis) | ± 200 g (3 tri-axis) | ± 200g (1 tri-axis) |
| **Measurement range (Gyro)** | ± 40 rad/s (1 tri-axis) | ± 70 rad/s (1 tri-axis) | ± 35 rad/s (1 tri-axis) | ± 35 rad/s (1 tri-axis) | ± 35[a,c] rad/s ± 40[b] rad/s (1 tri-axis) |
| **Output time window** | 100 [-25, 75] ms | 200 [-49, 150] ms | 50 [0, 50] ms | 100 [-20, 80] ms | 60 [-15, 45] ms |
| **Triggering sensor** | Accel. | Accel. | Accel. | Accel. | Accel. |
| **Triggering threshold** | 10 G | 10 G | 5 G | 8 G | 5 G |
| **No. measured impacts** | 1 | 771 | 2280 | 795 | 5 |

*a: Gymnastics, b: ice hockey, c: American football*

**Supplementary Data Table 8.** Results of Kolmogorov Smirnov test to compare the distributions of measured kinematics between mouthguard devices. Distributions for measured peak linear acceleration (g), peak rotational velocity (rad/s), and their primary frequencies (Hz) were compared.

| **Mouthguard 1** | **Mouthguard 2** | **P value** | | | |
|---|---|---|---|---|---|
| | | **a** (g) | **a** (Hz) | **ω** (rad/s) | **ω** (Hz) |
| HITIQ | Stanford MiG | $3.42 \times 10^{-23}$ | $1.33 \times 10^{-318}$ | 0.020 | $2.45 \times 10^{-245}$ |
| HITIQ | Prevent | $9.78 \times 10^{-6}$ | $8.39 \times 10^{-223}$ | 0.002 | 0 |
| HITIQ | Wake Forest | 0.00713 | 0.00144 | 0.464 | $5.61 \times 10^{-4}$ |
| Stanford MiG | Prevent | $1.48 \times 10^{-18}$ | 0 | 0.922 | 0 |
| Stanford MiG | Wake Forest | 0.00940 | $2.80 \times 10^{-5}$ | 0.359 | $1.85 \times 10^{-3}$ |
| Prevent | Wake Forest | 0.00905 | $7.01 \times 10^{-4}$ | 0.374 | $1.93 \times 10^{-5}$ |

**Supplementary Data Table 9.** Results from LASSO-penalized logistic regression and dominance analysis for unique predictors screened for injury risk function development.

| **Predictor** | **General Dominance (McFadden $R^2$)** | **LASSO Coefficient (standardized)** |
|---|---|---|
| **a** | 0.108 | 0.912 |
| **ω** | 0.065 | 0.842 |
| **α** | 0.061 | 0.380 |
| **HIC36** | 0.051 | –0.097 |
| **MPS95** | 0.050 | 0.206 |
| **$HIP_{rot}$** | 0.046 | –0.005 |
| **BrIC** | 0.046 | –0.362 |
| **HIP** | 0.042 | 0.000 |
| **$HIP_{lin}$** | 0.041 | 0.000 |
| **CP** | 0.029 | –0.242 |

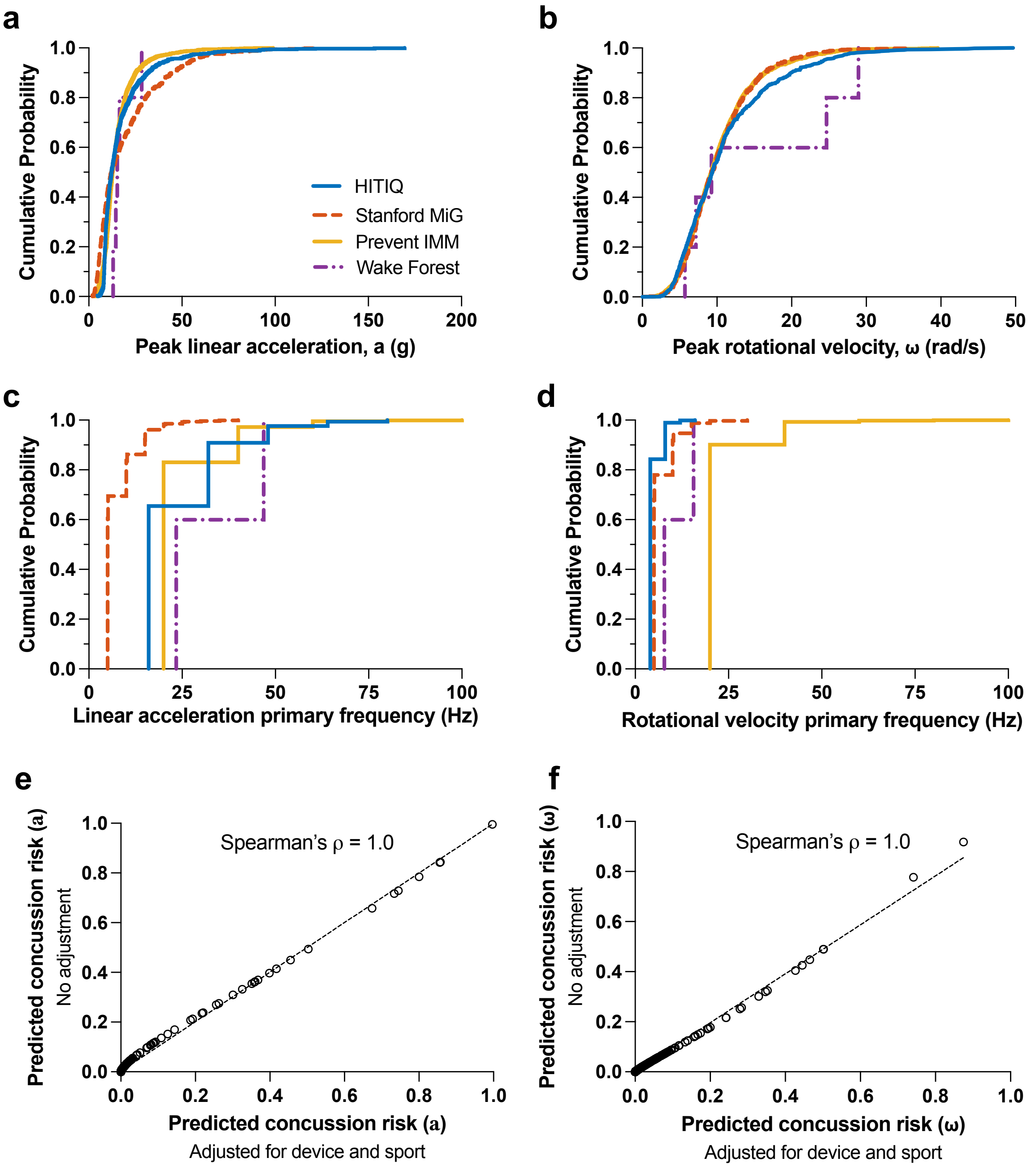

**Supplementary Data Figure 1.** Empirical cumulative distribution functions comparing peak kinematic measurements in the time and frequency domains across datasets with different mouthguard devices. **a,** Peak linear acceleration and **b,** peak rotational velocity distributions for each iMG. Distribution of primary frequencies for **c**, peak linear acceleration and **d**, peak rotational velocity. We note that HITIQ data reflect raw signals reprocessed using a 200 Hz low-pass filter, not the manufacturer's default output. **e-f,** Comparison of concussion risk predicted with and without adjusting the logistic regression model for mouthguard device type and sport using **e**, peak linear acceleration and **f**, peak rotational velocity. Despite measurable differences in kinematic distributions across mouthguard devices, adjusting for device and sport had minimal influence on predicted risk.

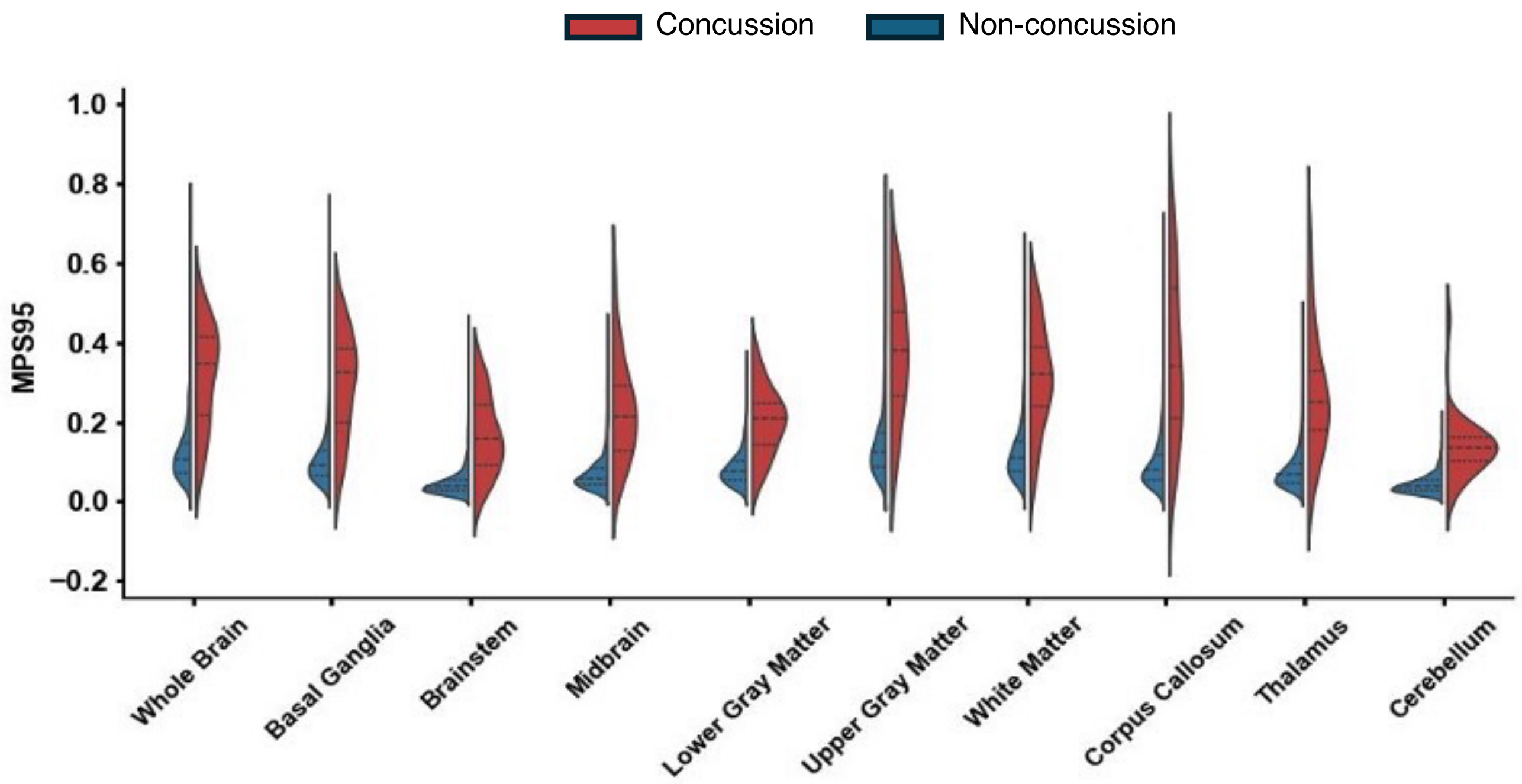

**Supplementary Data Figure 2.** Distribution of regional MPS95. Points are plotted per-player per-region. Dashed lines represent the median and interquartile rang

## Supplementary Data References

## Supplementary Information

### Comparison of mouthguard measurement devices

To assess measurement consistency across iMGs after 200 Hz filtering, we computed empirical cumulative distribution functions of peak linear acceleration ($\mathbf{a}$), peak rotational velocity ($\boldsymbol{\omega}$), and their primary frequencies for impacts recorded by each iMG (Supplementary Data Figure 1). These metrics were chosen because they are directly measured by the accelerometer(s) and gyroscope contained on each iMG sensor board. To characterize the primary frequency, we first computed the resultant of the triaxial measurements for linear acceleration and rotational velocity using the Euclidean norm. Each resultant trace was demeaned to remove any DC offset. To extract the primary frequency, we computed the discrete Fourier transform (`fft` in MATLAB) of each demeaned signal to obtain the single-sided amplitude spectrum. The frequency bin with the largest magnitude, excluding the zero-Hz component, was recorded as the primary frequency and used to construct cumulative distribution functions (Supplementary Data Figure 1c-d). In parallel, we extracted the peak values of the resultant $\mathbf{a}$ and $\boldsymbol{\omega}$ time traces and plotted their cumulative distribution functions (Supplementary Data Figure 1a-b).

To quantitatively assess whether the distributions differed significantly across iMGs, we performed pairwise two-sample Kolmogorov–Smirnov (KS) tests for each metric. These revealed statistically significant differences ($p < 0.05$) in both time- and frequency-domain metrics between several device pairs (Supplementary Data Table 8). While KS tests can detect differences in measurement distributions, they do not necessarily indicate that these differences meaningfully alter injury risk predictions. To evaluate this, we constructed two Bayesian logistic regression models for $\mathbf{a}$ and $\boldsymbol{\omega}$, including random intercepts for mouthguard type and sport, as well as an offset to account for the rare incidence of concussion (5.5 concussions per 1000 impacts).[1] We compared predicted concussion probabilities from this model to those from our main model that did not include random effects. The correlation between predicted risks from the two models was extremely high across all predictors (Supplementary Data Figure 1), suggesting that despite distributional differences between devices, the influence on estimated concussion risk was negligible.

### Brain sensitivity to rotational velocity and acceleration

To evaluate whether brain strains in our dataset were more sensitive to peak rotational velocity ($\boldsymbol{\omega}$) or peak rotational acceleration ($\boldsymbol{\alpha}$), we plotted the 95$^{th}$ percentile maximum principal strain from the DAMAGE[2] model as a function of both $\boldsymbol{\omega}$ and $\boldsymbol{\alpha}$ and examined where human impacts fell relative to a fixed slope representing the brain's natural frequency (Supplementary Data Figure 3). First, we defined a two-dimensional grid of peak rotational velocities ($\boldsymbol{\omega}$: 0.1–100 rad/s) and peak rotational accelerations ($\boldsymbol{\alpha}$: 1–20,000 rad/s²) spanning the full range of head impacts in our dataset. For each ($\boldsymbol{\omega}, \boldsymbol{\alpha}$) pair, a half-sine rotational-acceleration pulse was constructed at 1 ms resolution over a 2 s window, and used as input to the DAMAGE model. The resulting triplets [$\boldsymbol{\omega}$, $\boldsymbol{\alpha}$, DAMAGE] formed a simulation-only design matrix. To create continuous contour surfaces, these discrete DAMAGE values were interpolated onto a regular ($\boldsymbol{\omega},\boldsymbol{\alpha}$) mesh using MATLAB's `griddata` function with a natural neighbor scheme. Next, we computed DAMAGE for the 3,852 human impacts in our dataset and overlaid them on the interpolated contours. Finally, to demarcate the kinematic boundary where $\boldsymbol{\omega}$ and $\boldsymbol{\alpha}$ contributions to brain strain are equal, we overlaid a straight reference line whose slope is directly proportional to the brain's natural frequency ($f_n$) using Equation 1,

$$\alpha = \pi f_n \omega \tag{1}$$

where $f_n$ was estimated to be 25.1 Hz across the coronal, axial, and horizontal anatomical planes.[3]

**Candidate models for classification and injury risk function development**

We evaluated twelve logistic regression models to compare the predictive value of linear, rotational, and strain-based metrics for concussion classification and injury risk modeling. These models were selected based on their prevalence in head injury biomechanics research and safety standards. To prioritize which models were included in injury risk function development and comparative statistical testing, we applied multiple screening approaches including deviance, Bayesian Information Criterion, LASSO-penalized logistic regression, and dominance analysis.

Deviance and Bayesian Information Criterion assessed model fit across the twelve candidates. LASSO and dominance analysis were conducted using the ten unique predictors represented in these models (Supplementary Data Table 9). LASSO was performed using the `glmnet` R package with a binomial family and 5-fold cross-validation to identify the optimal penalty parameter, lambda. The L1 regularization penalty constrains the sum of the absolute values of the coefficients (n=10 coefficients) to be less than or equal to the specified constant determined by lambda, encouraging sparsity. Predictors were standardized prior to fitting. Dominance analysis was conducted on a full logistic regression model (n=10 predictors) using the `dominanceanalysis` R package. General dominance scores were computed using McFadden's pseudo-$R^2$ to quantify each predictor's average contribution across all possible subset models.

LASSO and dominance analysis also complemented our other statistical approaches. For instance, odds ratios quantified the direction and strength of each predictor's independent association with injury risk; area under the precision−recall curve measured the classification performance of each model in the context of class imbalance; LASSO identified relative predictor importance under regularized logistic regression; and dominance analysis estimated each predictor's average contribution to model fit. Across these approaches, $\mathbf{a}$, followed by $\boldsymbol{\omega}$ and $\boldsymbol{\alpha}$ consistently ranked among the most important and best-fitting predictor variables. As a result, we focused injury risk function development on models containing these three predictors. This prioritization ensured that the resulting injury risk functions were both well-fitting and practically useful, and reduced the number of formal statistical comparisons needed, thereby preserving statistical power on our limited concussion dataset. Other potential predictors, such as region-specific brain strains or directional kinematics, were excluded due to limited sample size of the concussions. We also excluded peak linear velocity, as it is not a widely used injury metric in safety standards or head injury research.

## References for Supplementary Information